# Dynamic backstepping control for pure-feedback nonlinear systems

ZHANG Sheng[*], QIAN Wei-qi

(2017.6)

Computational Aerodynamics Institution, China Aerodynamics Research and Development Center, Mianyang, 621000, China.
[*]zszhangshengzs@hotmail.com

**Abstract:** A dynamic backstepping method is proposed to design controllers for nonlinear systems in the pure-feedback form, for which the traditional backstepping method suffers from solving the implicit nonlinear algebraic equation. The idea of this method is to augment the (virtual) controls as states during each recursive step. As new dynamics are included in the design, the resulting controller is in the dynamic feedback form. Procedure of deriving the controller is detailed, and one more Lyapunov design is executed in each step compared with the traditional backstepping method. Under appropriate assumptions, the proposed control scheme achieves the uniformly asymptotically stability. The effectiveness of this method is illustrated by the stabilization and tracking numerical examples.

## 1. Introduction

The backstepping controller design methodology provides an effective tool of designing controllers for a large class of nonlinear systems with a triangular structure. Krstic, Kanellakopoulos, and Kokotovic [1] systematically developed this approach, from considering the exact model to encompassing the uncertain bounded nonlinearities and parameterized uncertainties. The basic idea behind backstepping is to break a design problem on the full system down to a sequence of sub-problems on lower order systems, and recursively use some states as "virtual controls" to obtain the intermediate control laws with the Control Lyapunov Function (CLF). Starting from lower order system and dealing with the interaction after augmentation of new dynamics make the controller design easy. The advantages of backstepping control include the guaranteed global of regional stability, the stress on robustness, and computable transient performance.

The backstepping method has received a great deal of interest since its proposition, and has been widely applied to the control problems arising from the aerospace engineering [2][3][4], mechanical engineering [5][6], etc. Along with these years of studies, this method has been evolved to be fairly systematical and inclusive. For example, techniques like the nonlinear damping [1], the variable structure control [7], the neural network adaptive control [8], and the fuzzy adaptive control [9] are synthesized to address various uncertainties, including the matching and un-matching. To resolve the problem of "explosion of terms", the dynamics surface control [10] and the constrained backstepping control [3] are further established. To address the deficiency of state information, the output feedback backstepping control is developed [1]. For the problem of control saturation, the limiting filters [3] and the boundedness propagation [11] are employed in the recursive design.

Nonetheless, within the extensive researches of backstepping method, the plants studied are usually in the form of strict-feedback. For the more general pure-feedback plants, which have no affine appearance of state variables to be used as virtual controls and the actual control, its usage may be restricted because the intractable implicit algebraic equations are encountered. Kanellakopoulos et al. [12] studied the adaptive control of parametric pure-feedback systems in a special form. Ge and Wang [13] used the neural networks to approximate the (virtual) controls out of the implicit algebraic equations, and proved that the control error will be ultimately bounded. Wang et al. [14] used the similar strategy and employed the input-to-state stability analysis and small-gain theorem to solve the "circularity problem" airing from the general pure-feedback problems. For a special class of pure-feedback system, Zou and Hou [15] employed filtered signals to circumvent algebraic loop problems and applied the compensator to counteract the resulting approximation errors. Wang et al. [16] exploited the mean value theorem to deduce the affine form of the pure-feedback plant to design the adaptive backstepping controller. In this paper, we will solve the algebraic loop problem from a different view. A dynamic backstepping method with stativization of (virtual) control is proposed. It circumvents the implicit algebraic equations and is widely applicable to the general pure-feedback nonlinear systems.

## 2. Basic backstepping method

We first briefly review the basic backstepping controller theory, and show its deficiency in treating pure-feedback plants. In introducing the method, the control objective is to stabilize the states of system towards the origin, which is assumed to be the equilibrium point. For other equilibrium points under the set-point control problems, with coordinate transformation they can be placed at the origin. Moreover, since the controller design procedure is similar for the high cascade model, for the sake of brevity, we only investigate the model of two cascades. For the strict-feedback plant described as

$$\dot{x}_1 = f_1(x_1) + g_1(x_1)x_2 \tag{1}$$

$$\dot{x}_2 = f_2(x_1, x_2) + g_2(x_1, x_2)u \tag{2}$$

where $x_1, x_2 \in \mathbb{R}^m$ are the states, $u \in \mathbb{R}^m$ is the control, $f_1: \mathbb{R}^m \to \mathbb{R}^m$, $f_2: \mathbb{R}^m \times \mathbb{R}^m \to \mathbb{R}^m$ are smooth nonlinear vector fields, and the matrix-valued functions $g_1: \mathbb{R}^m \to \mathbb{R}^{m \times m}$, $g_2: \mathbb{R}^m \times \mathbb{R}^m \to \mathbb{R}^{m \times m}$ are smooth and have inverse. The basic backstepping design procedure is summarised as follows.

**Step 1:** Consider Eq. (1). Regard $x_2$ as the virtual control in this equation and denote it with $x_{2d}$. The Lyapunov design is carried out as

CLF: $V_1 = \dfrac{1}{2}\boldsymbol{x}_1^{\mathrm{T}}\boldsymbol{x}_1$

$\Downarrow$

Drving $V_1$ decrease: $\dot{V}_1 = \boldsymbol{x}_1^{\mathrm{T}}\left(\boldsymbol{f}_1(\boldsymbol{x}_1) + \boldsymbol{g}_1(\boldsymbol{x}_1)\boldsymbol{x}_{2d}\right) \le 0$

$\Downarrow$ (3)

Algebraic control equation: $\boldsymbol{f}_1(\boldsymbol{x}_1) + \boldsymbol{g}_1(\boldsymbol{x}_1)\boldsymbol{x}_{2d} = \boldsymbol{\kappa}_1(\boldsymbol{x}_1)$

$\Downarrow$

Explicit virtual control expression: $\boldsymbol{x}_{2d} = \boldsymbol{g}_1^{-1}(-\boldsymbol{f}_1 + \boldsymbol{\kappa}_1(\boldsymbol{x}_1))$

where $\boldsymbol{\kappa}_1(\boldsymbol{x}_1)$ is the expected dynamics that satisfies the following two conditions: i) it drives $\dot{V}_1 < 0$ when $\boldsymbol{x}_1 \ne \boldsymbol{0}$, and ii) the resulting virtual control solution is continuous and $\left(\boldsymbol{g}_1(\boldsymbol{0})\right)^{-1}\left(-\boldsymbol{f}_1(\boldsymbol{0}) + \boldsymbol{\kappa}_1(\boldsymbol{0})\right) = \boldsymbol{0}$. Usually it may be set that $\boldsymbol{\kappa}_1(\boldsymbol{x}_1) = -\boldsymbol{K}_1\boldsymbol{x}_1$, where $\boldsymbol{K}_1$ is a positive gain matrix.

**Step 2:** Consider Eqs. (1) and (2) together, and construct a synthetic CLF to obtain

CLF: $V_2 = \dfrac{1}{2}\boldsymbol{x}_1^{\mathrm{T}}\boldsymbol{x}_1 + \dfrac{1}{2}(\boldsymbol{x}_2 - \boldsymbol{x}_{2d})^{\mathrm{T}}(\boldsymbol{x}_2 - \boldsymbol{x}_{2d})$

$\Downarrow$

Drving $V_1$ decrease: $\dot{V}_2 = \boldsymbol{x}_1^{\mathrm{T}}\left(\boldsymbol{f}_1(\boldsymbol{x}_1) + \boldsymbol{g}_1(\boldsymbol{x}_1)\boldsymbol{x}_2\right) + (\boldsymbol{x}_2 - \boldsymbol{x}_{2d})^{\mathrm{T}}(\dot{\boldsymbol{x}}_2 - \dot{\boldsymbol{x}}_{2d})$

$= \boldsymbol{x}_1^{\mathrm{T}}\left(\boldsymbol{f}_1(\boldsymbol{x}_1) + \boldsymbol{g}_1(\boldsymbol{x}_1)\boldsymbol{x}_{2d}\right) + (\boldsymbol{x}_2 - \boldsymbol{x}_{2d})^{\mathrm{T}}(\dot{\boldsymbol{x}}_2 - \dot{\boldsymbol{x}}_{2d} + \boldsymbol{g}_1(\boldsymbol{x}_1)^{\mathrm{T}}\boldsymbol{x}_1)$

$\le 0$ (4)

$\Downarrow$

Algebraic control equation: $\boldsymbol{f}_2(\boldsymbol{x}_1,\boldsymbol{x}_2) + \boldsymbol{g}_2(\boldsymbol{x}_1,\boldsymbol{x}_2)\boldsymbol{u} = \boldsymbol{\kappa}_2(\boldsymbol{x}_1,\boldsymbol{x}_2)$

$\Downarrow$

Explicit control expression: $\boldsymbol{u} = \boldsymbol{g}_2^{-1}\left(-\boldsymbol{f}_2 + \boldsymbol{\gamma}_2(\boldsymbol{x}_1,\boldsymbol{x}_2)\right)$

where $\boldsymbol{\kappa}_2(\boldsymbol{x}_1,\boldsymbol{x}_2)$ is the expected dynamics that driving $\dot{V}_2 < 0$ when $\boldsymbol{x}_2 \ne \boldsymbol{x}_{2d}$, and a common choice is that $\boldsymbol{\kappa}_2(\boldsymbol{x}_1,\boldsymbol{x}_2) = -\boldsymbol{K}_2(\boldsymbol{x}_2 - \boldsymbol{x}_{2d}) - \boldsymbol{g}_1(\boldsymbol{x}_1)^{\mathrm{T}}\boldsymbol{x}_1 + \dot{\boldsymbol{x}}_{2d}$, where $\boldsymbol{K}_2$ is a positive gain matrix. The resulting controller law $\boldsymbol{u}$ achieves the asymptotically stabilization since $\dot{V}_2 < 0$ except $\begin{bmatrix} \boldsymbol{x}_1 \\ \boldsymbol{x}_2 \end{bmatrix} = \boldsymbol{0}$.

Now consider the plant in the pure-feedback form like

$$\dot{\boldsymbol{x}}_1 = \boldsymbol{f}_1(\boldsymbol{x}_1,\boldsymbol{x}_2) \tag{5}$$

$$\dot{\boldsymbol{x}}_2 = \boldsymbol{f}_2(\boldsymbol{x}_1,\boldsymbol{x}_2,\boldsymbol{u}) \tag{6}$$

where $\boldsymbol{x}_1,\boldsymbol{x}_2 \in \mathbb{R}^m$ are the states, $\boldsymbol{u} \in \mathbb{R}^m$ is the control. Proceeding accordingly as Step 1 we may obtain the algebraic control equation

$$f_1(x_1, x_{2d}) = \kappa_1(x_1) \tag{7}$$

However, different from the strict-feedback system that has a affine structure, for the pure-feedback system, it may be not possible to get the explicit expression of $x_{2d}$ due to the implicit nonlinear form of Eq. (7), and this confines the application of the traditional backstepping method.

## 3. Dynamic backstepping method

To gain a systematic solution for the controller design problems on pure-feedback systems, a dynamic backstepping method is proposed. We first present the stabilizing control law and then expound the procedure to help understanding.

### 3.1 Main results

Assumptions that make the results rigorous are first presented.

**Assumption 1:** The exact solution for the implicit nonlinear algebraic control equation exists, such as the existence of $x_{2d}$ that rigorously nullifying Eq. (7).

**Assumption 2:** For the controlled domain $\mathbb{D}$ that contains the origin, the Jacobi matrixes $\dfrac{\partial f_1}{\partial x_2}$, $\dfrac{\partial f_1}{\partial x_{2d}}$ and $\dfrac{\partial f_2}{\partial u}$ are invertible.

**Assumption 3:** For the controlled domain $\mathbb{D}$ that contains the origin, $f_1(x_1, a) \neq f_1(x_1, b)$ when $a \neq b$.

Assumption 1 guarantees the theoretic existence of controller law that stabilizes the closed-loop system under the frame of backstepping design. Assumption 2 is related to Assumption 1 with the implicit function theorem [17]. Also note that $x_{2d}$ is a augmented state in the proposed method, and $\dfrac{\partial f}{\partial x_{2d}}$ is invertible when $x_{2d}$ is close to $x_2$, which is described in the assumed domain $\mathbb{D}$. Assumption 3 is used to deduce a general result for the controlled plant in this section, and it may be removed under certain situations, which will be shown in Sec. 4.

**Theorem 1:** Consider the pure-feedback cascade plant of the form

$$\dot{x}_1 = f_1(x_1, x_2) \tag{8}$$

$$\dot{x}_2 = f_2(x_1, x_2, u) \tag{9}$$

where $x_1, x_2 \in \mathbb{R}^m$ are the states, $u \in \mathbb{R}^m$ is the control, and $f_i$ (i = 1, 2) are smooth vector fields. If Assumption 1 holds, then the following dynamic feedback control law stabilizes the plant in the domain $\mathbb{D}$, where Assumptions 2 and 3 hold.

$$u = \int_0^t \left\{ -K_{v2} \left( \frac{\partial h_2}{\partial u} \right)^{\mathrm{T}} h_2 - \left( \frac{\partial h_2}{\partial u} \right)^{-1} \left( \begin{array}{l} \frac{\partial h_2}{\partial x_1} f_1(x_1, x_2) + \frac{\partial h_2}{\partial x_2} f_2(x_1, x_2, u) + \frac{\partial h_2}{\partial x_{2d}} \dot{x}_{2d} \\ + \left( \frac{\partial f_1(x_1, x_2)}{\partial x_2} \right)^{\mathrm{T}} \left( f_1(x_1, x_2) - f_1(x_1, x_{2d}) \right) \end{array} \right) \right\} \mathrm{d}t, \left. u \right|_{t=0} = u_0 \quad (10)$$

where $x_{2d}$ is a augmented state and its dynamics is

$$x_{2d} = \int_0^t \left\{ -K_{v1} \left( \frac{\partial h_1}{\partial x_{2d}} \right)^{\mathrm{T}} h_1 - \left( \frac{\partial h_1}{\partial x_{2d}} \right)^{-1} \left( \frac{\partial h_1}{\partial x_1} f_1(x_1, x_{2d}) + x_1 \right) \right\} \mathrm{d}t, \left. x_{2d} \right|_{t=0} = x_{2d0} \quad (11)$$

$u_0$ and $x_{2d0}$ are the initial values for the augmented states. The other relevant quantities are

$$h_1(x_1, x_{2d}) = f_1(x_1, x_{2d}) - \kappa_1(x_1) \quad (12)$$

$$h_2(x_1, x_2, u, x_{2d}) = f_2(x_1, x_2, u) - \kappa_2(x_1, x_2, x_{2d}) \quad (13)$$

$\kappa_1(x_1)$ may be arbitrary function that satisfies: i) $x_1^{\mathrm{T}} \kappa_1(x_1) < 0$ when $x_1 \neq \boldsymbol{0}$, and ii) the mapping $x_{2d} = C_1(x_1)$ implicitly determined by Eq. (12) is continuous and $C_1(\boldsymbol{0}) = \boldsymbol{0}$.

$$\kappa_2(x_1, x_2, x_{2d}) = \Gamma(x_1, x_2, x_{2d}) - \left( \frac{\partial f_1(x_1, x_2)}{\partial x_2} \right)^{-1} \left( x_1 + \left( \frac{\partial h_1}{\partial x_1} \right)^{\mathrm{T}} h_1 + \left( \frac{\partial f_1(x_1, x_2)}{\partial x_1} - \frac{\partial f_1(x_1, x_{2d})}{\partial x_1} \right) f_1(x_1, x_2) - \frac{\partial f_1(x_1, x_{2d})}{\partial x_{2d}} \dot{x}_{2d} \right) \quad (14)$$

and $\Gamma(x_1, x_2, x_{2d})$ may be arbitrary function that satisfies $\left( f_1(x_1, x_2) - f_1(x_1, x_{2d}) \right)^{\mathrm{T}} \frac{\partial f_1(x_1, x_2)}{\partial x_2} \Gamma(x_1, x_2, x_{2d}) < 0$ when $x_2 \neq x_{2d}$.

### 3.2 Proof of Theorem 1

Construct a CLF as

$$V = \frac{1}{2} x_1^{\mathrm{T}} x_1 + \frac{1}{2} h_1^{\mathrm{T}} h_1 + \frac{1}{2} \left( f_1(x_1, x_2) - f_1(x_1, x_{2d}) \right)^{\mathrm{T}} \left( f_1(x_1, x_2) - f_1(x_1, x_{2d}) \right) + \frac{1}{2} h_2^{\mathrm{T}} h_2 \quad (15)$$

where $h_1$ and $h_2$ are given by Eqs. (12) and (13) respectively. Differentiating Eq. (15) renders

$$\dot{V} = x_1^{\mathrm{T}} \dot{x}_1 + h_1^{\mathrm{T}} \left( \frac{\partial h_1}{\partial x_1} \dot{x}_1 + \frac{\partial h_1}{\partial x_{2d}} \dot{x}_{2d} \right) + \left( f_1(x_1, x_2) - f_1(x_1, x_{2d}) \right)^{\mathrm{T}} \left( \dot{f}_1(x_1, x_2) - \dot{f}_1(x_1, x_{2d}) \right)$$

$$+ h_2^{\mathrm{T}} \left( \frac{\partial h_2}{\partial x_1} \dot{x}_1 + \frac{\partial h_2}{\partial x_2} \dot{x}_2 + \frac{\partial h_2}{\partial x_{2d}} \dot{x}_{2d} + \frac{\partial h_2}{\partial u} \dot{u} \right) \quad (16)$$

with certain treatment to deal with the interactions, there is

$$\dot{V} = \boldsymbol{x}_1{}^{\mathrm{T}}\boldsymbol{\kappa}_1(\boldsymbol{x}_1) + \boldsymbol{h}_1{}^{\mathrm{T}}\boldsymbol{x}_1 + \big(\boldsymbol{f}_1(\boldsymbol{x}_1,\boldsymbol{x}_2) - \boldsymbol{f}_1(\boldsymbol{x}_1,\boldsymbol{x}_{2d})\big)^{\mathrm{T}}\boldsymbol{x}_1$$

$$+ \boldsymbol{h}_1{}^{\mathrm{T}}\left(\frac{\partial \boldsymbol{h}_1}{\partial \boldsymbol{x}_1}\boldsymbol{f}_1(\boldsymbol{x}_1,\boldsymbol{x}_{2d}) + \frac{\partial \boldsymbol{h}_1}{\partial \boldsymbol{x}_{2d}}\dot{\boldsymbol{x}}_{2d}\right) + \big(\boldsymbol{f}_1(\boldsymbol{x}_1,\boldsymbol{x}_2) - \boldsymbol{f}_1(\boldsymbol{x}_1,\boldsymbol{x}_{2d})\big)^{\mathrm{T}}\left(\frac{\partial \boldsymbol{h}_1}{\partial \boldsymbol{x}_1}\right)^{\mathrm{T}}\boldsymbol{h}_1$$

$$+ \big(\boldsymbol{f}_1(\boldsymbol{x}_1,\boldsymbol{x}_2) - \boldsymbol{f}_1(\boldsymbol{x}_1,\boldsymbol{x}_{2d})\big)^{\mathrm{T}}\left(\frac{\partial \boldsymbol{f}_1(\boldsymbol{x}_1,\boldsymbol{x}_2)}{\partial \boldsymbol{x}_1} - \frac{\partial \boldsymbol{f}_1(\boldsymbol{x}_1,\boldsymbol{x}_{2d})}{\partial \boldsymbol{x}_1}\right)\boldsymbol{f}_1(\boldsymbol{x}_1,\boldsymbol{x}_2)$$

$$+ \big(\boldsymbol{f}_1(\boldsymbol{x}_1,\boldsymbol{x}_2) - \boldsymbol{f}_1(\boldsymbol{x}_1,\boldsymbol{x}_{2d})\big)^{\mathrm{T}}\left(\frac{\partial \boldsymbol{f}_1(\boldsymbol{x}_1,\boldsymbol{x}_2)}{\partial \boldsymbol{x}_2}\boldsymbol{\kappa}_2(\boldsymbol{x}_1,\boldsymbol{x}_2,\boldsymbol{x}_{2d}) - \frac{\partial \boldsymbol{f}_1(\boldsymbol{x}_1,\boldsymbol{x}_{2d})}{\partial \boldsymbol{x}_{2d}}\dot{\boldsymbol{x}}_{2d}\right) \tag{17}$$

$$+ \boldsymbol{h}_2{}^{\mathrm{T}}\left(\frac{\partial \boldsymbol{f}_1(\boldsymbol{x}_1,\boldsymbol{x}_2)}{\partial \boldsymbol{x}_2}\right)^{\mathrm{T}}\big(\boldsymbol{f}_1(\boldsymbol{x}_1,\boldsymbol{x}_2) - \boldsymbol{f}_1(\boldsymbol{x}_1,\boldsymbol{x}_{2d})\big)$$

$$+ \boldsymbol{h}_2{}^{\mathrm{T}}\left(\frac{\partial \boldsymbol{h}_2}{\partial \boldsymbol{x}_1}\boldsymbol{f}_1(\boldsymbol{x}_1,\boldsymbol{x}_2) + \frac{\partial \boldsymbol{h}_2}{\partial \boldsymbol{x}_2}\boldsymbol{f}_2(\boldsymbol{x}_1,\boldsymbol{x}_2,\boldsymbol{u}) + \frac{\partial \boldsymbol{h}_2}{\partial \boldsymbol{x}_{2d}}\dot{\boldsymbol{x}}_{2d} + \frac{\partial \boldsymbol{h}_2}{\partial \boldsymbol{u}}\dot{\boldsymbol{u}}\right)$$

Substituting the differential form of Eq. (11), i.e.

$$\dot{\boldsymbol{x}}_{2d} = -\boldsymbol{K}_{v1}\left(\frac{\partial \boldsymbol{f}_1(\boldsymbol{x}_1,\boldsymbol{x}_{2d})}{\partial \boldsymbol{x}_{2d}}\right)^{\mathrm{T}}\boldsymbol{h}_1 - \left(\frac{\partial \boldsymbol{f}_1(\boldsymbol{x}_1,\boldsymbol{x}_{2d})}{\partial \boldsymbol{x}_{2d}}\right)^{-1}\left(\frac{\partial \boldsymbol{h}_1}{\partial \boldsymbol{x}_1}\boldsymbol{f}_1(\boldsymbol{x}_1,\boldsymbol{x}_{2d}) + \boldsymbol{x}_1\right) \tag{18}$$

we obtain

$$\dot{V} = \boldsymbol{x}_1{}^{\mathrm{T}}\boldsymbol{\kappa}_1(\boldsymbol{x}_1) - \boldsymbol{h}_1{}^{\mathrm{T}}\frac{\partial \boldsymbol{f}_1(\boldsymbol{x}_1,\boldsymbol{x}_{2d})}{\partial \boldsymbol{x}_{2d}}\boldsymbol{K}_{v1}\left(\frac{\partial \boldsymbol{f}_1(\boldsymbol{x}_1,\boldsymbol{x}_{2d})}{\partial \boldsymbol{x}_{2d}}\right)^{\mathrm{T}}\boldsymbol{h}_1$$

$$+ \big(\boldsymbol{f}_1(\boldsymbol{x}_1,\boldsymbol{x}_2) - \boldsymbol{f}_1(\boldsymbol{x}_1,\boldsymbol{x}_{2d})\big)^{\mathrm{T}}\left(\boldsymbol{x}_1 + \left(\frac{\partial \boldsymbol{h}_1}{\partial \boldsymbol{x}_1}\right)^{\mathrm{T}}\boldsymbol{h}_1\right)$$

$$+ \big(\boldsymbol{f}_1(\boldsymbol{x}_1,\boldsymbol{x}_2) - \boldsymbol{f}_1(\boldsymbol{x}_1,\boldsymbol{x}_{2d})\big)^{\mathrm{T}}\left(\frac{\partial \boldsymbol{f}_1(\boldsymbol{x}_1,\boldsymbol{x}_2)}{\partial \boldsymbol{x}_1} - \frac{\partial \boldsymbol{f}_1(\boldsymbol{x}_1,\boldsymbol{x}_{2d})}{\partial \boldsymbol{x}_1}\right)\boldsymbol{f}_1(\boldsymbol{x}_1,\boldsymbol{x}_2)$$

$$+ \big(\boldsymbol{f}_1(\boldsymbol{x}_1,\boldsymbol{x}_2) - \boldsymbol{f}_1(\boldsymbol{x}_1,\boldsymbol{x}_{2d})\big)^{\mathrm{T}}\left(\frac{\partial \boldsymbol{f}_1(\boldsymbol{x}_1,\boldsymbol{x}_2)}{\partial \boldsymbol{x}_2}\boldsymbol{\kappa}_2(\boldsymbol{x}_1,\boldsymbol{x}_2,\boldsymbol{x}_{2d}) - \frac{\partial \boldsymbol{f}_1(\boldsymbol{x}_1,\boldsymbol{x}_{2d})}{\partial \boldsymbol{x}_{2d}}\dot{\boldsymbol{x}}_{2d}\right) \tag{19}$$

$$+ \boldsymbol{h}_2{}^{\mathrm{T}}\left(\frac{\partial \boldsymbol{f}_1(\boldsymbol{x}_1,\boldsymbol{x}_2)}{\partial \boldsymbol{x}_2}\right)^{\mathrm{T}}\big(\boldsymbol{f}_1(\boldsymbol{x}_1,\boldsymbol{x}_2) - \boldsymbol{f}_1(\boldsymbol{x}_1,\boldsymbol{x}_{2d})\big)$$

$$+ \boldsymbol{h}_2{}^{\mathrm{T}}\left(\frac{\partial \boldsymbol{h}_2}{\partial \boldsymbol{x}_1}\boldsymbol{f}_1(\boldsymbol{x}_1,\boldsymbol{x}_2) + \frac{\partial \boldsymbol{h}_2}{\partial \boldsymbol{x}_2}\boldsymbol{f}_2(\boldsymbol{x}_1,\boldsymbol{x}_2,\boldsymbol{u}) + \frac{\partial \boldsymbol{h}_2}{\partial \boldsymbol{x}_{2d}}\dot{\boldsymbol{x}}_{2d} + \frac{\partial \boldsymbol{h}_2}{\partial \boldsymbol{u}}\dot{\boldsymbol{u}}\right)$$

Substituting Eq. (14) and the differential form of Eq. (10), i.e.

$$\dot{\boldsymbol{u}} = -\boldsymbol{K}_{v2}\left(\frac{\partial \boldsymbol{h}_2}{\partial \boldsymbol{u}}\right)^{\mathrm{T}}\boldsymbol{h}_2 - \left(\frac{\partial \boldsymbol{h}_2}{\partial \boldsymbol{u}}\right)^{-1}\left(\begin{array}{l}\dfrac{\partial \boldsymbol{h}_2}{\partial \boldsymbol{x}_1}\boldsymbol{f}_1(\boldsymbol{x}_1,\boldsymbol{x}_2) + \dfrac{\partial \boldsymbol{h}_2}{\partial \boldsymbol{x}_2}\boldsymbol{f}_2(\boldsymbol{x}_1,\boldsymbol{x}_2,\boldsymbol{u}) + \dfrac{\partial \boldsymbol{h}_2}{\partial \boldsymbol{x}_{2d}}\dot{\boldsymbol{x}}_{2d} \\ + \left(\dfrac{\partial \boldsymbol{f}_1(\boldsymbol{x}_1,\boldsymbol{x}_2)}{\partial \boldsymbol{x}_2}\right)^{\mathrm{T}}\big(\boldsymbol{f}_1(\boldsymbol{x}_1,\boldsymbol{x}_2) - \boldsymbol{f}_1(\boldsymbol{x}_1,\boldsymbol{x}_{2d})\big)\end{array}\right) \tag{20}$$

we may further obtain that

$$\dot{V} = \begin{pmatrix} \boldsymbol{x}_1{}^{\mathrm{T}}\boldsymbol{\kappa}_1(\boldsymbol{x}_1) - \boldsymbol{h}_1{}^{\mathrm{T}}\dfrac{\partial \boldsymbol{f}_1(\boldsymbol{x}_1,\boldsymbol{x}_{2d})}{\partial \boldsymbol{x}_{2d}}\boldsymbol{K}_{v1}\left(\dfrac{\partial \boldsymbol{f}_1(\boldsymbol{x}_1,\boldsymbol{x}_{2d})}{\partial \boldsymbol{x}_{2d}}\right)^{\mathrm{T}}\boldsymbol{h}_1 \\[2ex] \big(\boldsymbol{f}_1(\boldsymbol{x}_1,\boldsymbol{x}_2) - \boldsymbol{f}_1(\boldsymbol{x}_1,\boldsymbol{x}_{2d})\big)^{\mathrm{T}}\dfrac{\partial \boldsymbol{f}_1(\boldsymbol{x}_1,\boldsymbol{x}_2)}{\partial \boldsymbol{x}_2}\boldsymbol{\Gamma}(\boldsymbol{x}_1,\boldsymbol{x}_2,\boldsymbol{x}_{2d}) \\[2ex] -\boldsymbol{h}_2{}^{\mathrm{T}}\dfrac{\partial \boldsymbol{f}_2(\boldsymbol{x}_1,\boldsymbol{x}_2,\boldsymbol{u})}{\partial \boldsymbol{u}}\boldsymbol{K}_{v2}\left(\dfrac{\partial \boldsymbol{f}_2(\boldsymbol{x}_1,\boldsymbol{x}_2,\boldsymbol{u})}{\partial \boldsymbol{u}}\right)^{\mathrm{T}}\boldsymbol{h}_2 \end{pmatrix} \le 0 \tag{21}$$

Since $\dot{V} < 0$ for $\mathbb{D}\,/\,\{\boldsymbol{0}\}$, this proves that the control law given by Eq. (10) will asymptotically stabilize the plant described by Eqs. (8) and (9).

### 3.3 Design procedure

Readers may be enlightened by the proof about the controller design, yet to facilitate understanding how the control law are constructed, the concrete procedure will be presented.

**Step 1:** Consider Eq. (8) only. In this equation regard $\boldsymbol{x}_2$ as the virtual control and denote it with $\boldsymbol{x}_{2d}$, a CLF is constructed to obtain the algebraic control equation

$$V_{11} = \frac{1}{2}\boldsymbol{x}_1{}^{\mathrm{T}}\boldsymbol{x}_1$$
$$\Downarrow$$
$$\dot{V}_{11} = \boldsymbol{x}_1{}^{\mathrm{T}}\boldsymbol{f}_1(\boldsymbol{x}_1,\boldsymbol{x}_{2d}) \le 0 \tag{22}$$
$$\Downarrow$$
$$\boldsymbol{f}_1(\boldsymbol{x}_1,\boldsymbol{x}_{2d}) = \boldsymbol{\kappa}_1(\boldsymbol{x}_1)$$

where $\boldsymbol{\kappa}_1(\boldsymbol{x}_1)$ satisfies the conditions in Theorem 1, and generally it may be set to be $\boldsymbol{\kappa}_1(\boldsymbol{x}_1) = -\boldsymbol{K}_1\boldsymbol{x}_1$, where $\boldsymbol{K}_1$ is a positive gain matrix. Since we may not be able to obtain the analytic expression of $\boldsymbol{x}_{2d}{}^* = \boldsymbol{C}_1(\boldsymbol{x}_1)$ that rigorously satisfy $\boldsymbol{f}_1(\boldsymbol{x}_1,\boldsymbol{x}_{2d}{}^*) - \boldsymbol{\kappa}_1(\boldsymbol{x}_1) \equiv \boldsymbol{0}$, we circumvent such problem by considering the dynamics of the virtual control $\boldsymbol{x}_{2d}$, in the hope that $\boldsymbol{x}_{2d}$ will satisfy the implicit algebraic equation in a asymptotical way. To realize this, again a CLF is constructed as

$$V_{12} = V_{11} + \frac{1}{2}\boldsymbol{h}_1{}^{\mathrm{T}}\boldsymbol{h}_1 \tag{23}$$

where $\boldsymbol{h}_1$ is given by Eq. (12). Differentiating $V_{12}$ and driving $\dot{V}_{12} \le 0$, we have

$$\dot{V}_{12} = \boldsymbol{x}_1^{\mathrm{T}} \boldsymbol{f}_1(\boldsymbol{x}_1, \boldsymbol{x}_{2d}) + \boldsymbol{h}_1^{\mathrm{T}} \left( \frac{\partial \boldsymbol{h}_1}{\partial \boldsymbol{x}_1} \dot{\boldsymbol{x}}_1 + \frac{\partial \boldsymbol{h}_1}{\partial \boldsymbol{x}_{2d}} \dot{\boldsymbol{x}}_{2d} \right)$$

$$= \boldsymbol{x}_1^{\mathrm{T}} \boldsymbol{\kappa}_1(\boldsymbol{x}_1) + \boldsymbol{h}_1^{\mathrm{T}} \left( \boldsymbol{x}_1 + \frac{\partial \boldsymbol{h}_1}{\partial \boldsymbol{x}_1} \dot{\boldsymbol{x}}_1 + \frac{\partial \boldsymbol{h}_1}{\partial \boldsymbol{x}_{2d}} \dot{\boldsymbol{x}}_{2d} \right) \le 0 \tag{24}$$

$$\Downarrow$$

$$\dot{\boldsymbol{x}}_{2d} = -\boldsymbol{K}_{v1} \left( \frac{\partial \boldsymbol{h}_1}{\partial \boldsymbol{x}_{2d}} \right)^{\mathrm{T}} \boldsymbol{h}_1 - \left( \frac{\partial \boldsymbol{h}_1}{\partial \boldsymbol{x}_{2d}} \right)^{-1} \left( \frac{\partial \boldsymbol{h}_1}{\partial \boldsymbol{x}_1} \boldsymbol{f}_1(\boldsymbol{x}_1, \boldsymbol{x}_{2d}) + \boldsymbol{x}_1 \right)$$

Thus we have the virtual control $\boldsymbol{x}_{2d}$ that achieves $\lim\limits_{t \to \infty} \boldsymbol{h}_1 = 0$ and $\lim\limits_{t \to \infty} \boldsymbol{x}_1 = 0$. For the applicability of the equivalent integral form given by Eq. (11), we use Assumptions 1 and 2 to guarantee that $\boldsymbol{x}_{2d}*$ exists and the initial value $\boldsymbol{x}_{2d0}$ is close to $\boldsymbol{x}_{2d}*(t_0) = \boldsymbol{C}_1\big(\boldsymbol{x}_1(t_0)\big)$ where $\left( \frac{\partial \boldsymbol{f}_1}{\partial \boldsymbol{x}_{2d}} \right)^{-1}$, i.e., $\left( \frac{\partial \boldsymbol{h}_1}{\partial \boldsymbol{x}_{2d}} \right)^{-1}$ exists.

**Step 2:** Consider Eqs. (8) and (9) together, and construct a CLF which aims to track $\boldsymbol{x}_{2d}$ in virtue of Assumption 3

$$V_{21} = V_{12} + \frac{1}{2} \big( \boldsymbol{f}_1(\boldsymbol{x}_1, \boldsymbol{x}_2) - \boldsymbol{f}_1(\boldsymbol{x}_1, \boldsymbol{x}_{2d}) \big)^{\mathrm{T}} \big( \boldsymbol{f}_1(\boldsymbol{x}_1, \boldsymbol{x}_2) - \boldsymbol{f}_1(\boldsymbol{x}_1, \boldsymbol{x}_{2d}) \big) \tag{25}$$

Proceeding similarly through differentiating $V_{21}$ and driving $\dot{V}_{21} \le 0$ gives

$$\dot{V}_{21} = \boldsymbol{x}_1^{\mathrm{T}} \boldsymbol{f}_1(\boldsymbol{x}_1, \boldsymbol{x}_{2d}) + \big( \boldsymbol{f}_1(\boldsymbol{x}_1, \boldsymbol{x}_2) - \boldsymbol{f}_1(\boldsymbol{x}_1, \boldsymbol{x}_{2d}) \big)^{\mathrm{T}} \boldsymbol{x}_1$$

$$+ \boldsymbol{h}_1^{\mathrm{T}} \left( \frac{\partial \boldsymbol{h}_1}{\partial \boldsymbol{x}_1} \boldsymbol{f}_1(\boldsymbol{x}_1, \boldsymbol{x}_{2d}) + \frac{\partial \boldsymbol{h}_1}{\partial \boldsymbol{x}_{2d}} \dot{\boldsymbol{x}}_{2d} \right) + \big( \boldsymbol{f}_1(\boldsymbol{x}_1, \boldsymbol{x}_2) - \boldsymbol{f}_1(\boldsymbol{x}_1, \boldsymbol{x}_{2d}) \big)^{\mathrm{T}} \left( \frac{\partial \boldsymbol{h}_1}{\partial \boldsymbol{x}_1} \right)^{\mathrm{T}} \boldsymbol{h}_1$$

$$+ \big( \boldsymbol{f}_1(\boldsymbol{x}_1, \boldsymbol{x}_2) - \boldsymbol{f}_1(\boldsymbol{x}_1, \boldsymbol{x}_{2d}) \big)^{\mathrm{T}} \left( \frac{\partial \boldsymbol{f}_1(\boldsymbol{x}_1, \boldsymbol{x}_2)}{\partial \boldsymbol{x}_1} - \frac{\partial \boldsymbol{f}_1(\boldsymbol{x}_1, \boldsymbol{x}_{2d})}{\partial \boldsymbol{x}_1} \right) \boldsymbol{f}_1(\boldsymbol{x}_1, \boldsymbol{x}_2)$$

$$+ \big( \boldsymbol{f}_1(\boldsymbol{x}_1, \boldsymbol{x}_2) - \boldsymbol{f}_1(\boldsymbol{x}_1, \boldsymbol{x}_{2d}) \big)^{\mathrm{T}} \left( \frac{\partial \boldsymbol{f}_1(\boldsymbol{x}_1, \boldsymbol{x}_2)}{\partial \boldsymbol{x}_2} \dot{\boldsymbol{x}}_2 - \frac{\partial \boldsymbol{f}_1(\boldsymbol{x}_1, \boldsymbol{x}_{2d})}{\partial \boldsymbol{x}_{2d}} \dot{\boldsymbol{x}}_{2d} \right) \le 0 \tag{26}$$

$$\Downarrow$$

$$\boldsymbol{f}_2(\boldsymbol{x}_1, \boldsymbol{x}_2, \boldsymbol{u}) = \boldsymbol{\kappa}_2(\boldsymbol{x}_1, \boldsymbol{x}_2, \boldsymbol{x}_{2d})$$

where $\boldsymbol{\kappa}_2(\boldsymbol{x}_1, \boldsymbol{x}_2, \boldsymbol{x}_{2d})$ is given by Eq. (14), and a feasible form of $\boldsymbol{\Gamma}(\boldsymbol{x}_1, \boldsymbol{x}_2, \boldsymbol{x}_{2d})$ may be

$$\boldsymbol{\Gamma}(\boldsymbol{x}_1, \boldsymbol{x}_2, \boldsymbol{x}_{2d}) = -\boldsymbol{K}_2 \left( \frac{\partial \boldsymbol{f}_1(\boldsymbol{x}_1, \boldsymbol{x}_2)}{\partial \boldsymbol{x}_2} \right)^{\mathrm{T}} \big( \boldsymbol{f}_1(\boldsymbol{x}_1, \boldsymbol{x}_2) - \boldsymbol{f}_1(\boldsymbol{x}_1, \boldsymbol{x}_{2d}) \big) \tag{27}$$

where $\boldsymbol{K}_2$ is a positive gain matrix. Analogously, we consider the dynamics of the control $\boldsymbol{u}$ to avoid the difficulty in searching its analytic solution. The CLF for the whole plant is constructed as

$$V_{22} = V_{21} + \frac{1}{2}\boldsymbol{h}_2{}^{\mathrm{T}}\boldsymbol{h}_2 \tag{28}$$

where $\boldsymbol{h}_2$ is given by Eq. (13). $V_{22}$ is just the CLF constructed in the proof of Theorem 1. Then the controller that guarantee $V_{22}$ decreases is

$$\dot{V}_{22} \leq 0$$
$$\Downarrow \tag{29}$$
$$\dot{\boldsymbol{u}} = -\boldsymbol{K}_{v2}\left(\frac{\partial \boldsymbol{h}_2}{\partial \boldsymbol{u}}\right)^{\mathrm{T}}\boldsymbol{h}_2 - \left(\frac{\partial \boldsymbol{h}_2}{\partial \boldsymbol{u}}\right)^{-1}\left(\begin{array}{l}\dfrac{\partial \boldsymbol{h}_2}{\partial \boldsymbol{x}_1}\boldsymbol{f}_1(\boldsymbol{x}_1,\boldsymbol{x}_2) + \dfrac{\partial \boldsymbol{h}_2}{\partial \boldsymbol{x}_2}\boldsymbol{f}_2(\boldsymbol{x}_1,\boldsymbol{x}_2,\boldsymbol{u}) + \dfrac{\partial \boldsymbol{h}_2}{\partial \boldsymbol{x}_{2d}}\dot{\boldsymbol{x}}_{2d} \\ +\left(\dfrac{\partial \boldsymbol{f}_1(\boldsymbol{x}_1,\boldsymbol{x}_2)}{\partial \boldsymbol{x}_2}\right)^{\mathrm{T}}\left(\boldsymbol{f}_1(\boldsymbol{x}_1,\boldsymbol{x}_2) - \boldsymbol{f}_1(\boldsymbol{x}_1,\boldsymbol{x}_{2d})\right)\end{array}\right)$$

Also by Assumption 2, the existence of $\left(\dfrac{\partial \boldsymbol{h}_2}{\partial \boldsymbol{u}}\right)^{-1}$ is guaranteed. Thus, the equivalent integral form is obtained with a reasonable $\boldsymbol{u}_0$.

Through the design procedure, it is shown that there are two Lypunov designs during each step. This operation along with the augmentation of the dynamics of the (virtual) control is used to solve the implicit nonlinear algebraic control equations, i.e., $\boldsymbol{h}_1(\boldsymbol{x}_1,\boldsymbol{x}_{2d}) = \boldsymbol{0}$ and $\boldsymbol{h}_2(\boldsymbol{x}_1,\boldsymbol{x}_2,\boldsymbol{x}_{2d},\boldsymbol{u}) = \boldsymbol{0}$. With the dynamic feedback control law, the states and the implicit nonlinear algebraic equations are both driven to be zero.

## 4. Further discussion
### 4.1 Simplification of controller form

The controller design presented last section aims at the general situation, and the resulting controller is complex in its form. Simplification of the controller is possible upon some conditions on the control gains, and this is helpful to alleviate the problem of "explosion of terms".

In the second Lyapunov design of Step 1, presume that $\boldsymbol{\kappa}_1(\boldsymbol{x}_1) = -\boldsymbol{K}_1\boldsymbol{x}_1$, then with the condition that $\boldsymbol{K}_{v1}$ satisfies

$$\frac{\partial \boldsymbol{h}_1}{\partial \boldsymbol{x}_{2d}}\boldsymbol{K}_{v1}\left(\frac{\partial \boldsymbol{h}_1}{\partial \boldsymbol{x}_{2d}}\right)^{\mathrm{T}} > \frac{3}{4}\boldsymbol{K}_1{}^{-1} - \frac{1}{4}\left(\frac{\partial \boldsymbol{h}_1}{\partial \boldsymbol{x}_1} + (\frac{\partial \boldsymbol{h}_1}{\partial \boldsymbol{x}_1})^{\mathrm{T}}\right) + \frac{3}{4}\frac{\partial \boldsymbol{h}_1}{\partial \boldsymbol{x}_1}\boldsymbol{K}_1\left(\frac{\partial \boldsymbol{h}_1}{\partial \boldsymbol{x}_1}\right)^{\mathrm{T}} \tag{30}$$

the virtual control, i.e., $\boldsymbol{x}_{2d}$, may be set as

$$\dot{\boldsymbol{x}}_{2d} = -\boldsymbol{K}_{v1}\left(\frac{\partial \boldsymbol{h}_1}{\partial \boldsymbol{x}_{2d}}\right)^{\mathrm{T}}\boldsymbol{h}_1 \tag{31}$$

and it also guarantees the stability of the subsystem. This may be verified by substituting Eq. (31) into $\dot{V}_{12}$ to be

$$\dot{V}_{12} = \boldsymbol{x}_1^{\mathrm{T}} \boldsymbol{f}_1(\boldsymbol{x}_1, \boldsymbol{x}_{2d}) + \boldsymbol{h}_1^{\mathrm{T}} \left( \frac{\partial \boldsymbol{h}_1}{\partial \boldsymbol{x}_1} \dot{\boldsymbol{x}}_1 + \frac{\partial \boldsymbol{h}_1}{\partial \boldsymbol{x}_{2d}} \dot{\boldsymbol{x}}_{2d} \right)$$

$$= -\boldsymbol{x}_1^{\mathrm{T}} \boldsymbol{K}_1 \boldsymbol{x}_1 + \boldsymbol{h}_1^{\mathrm{T}} (\boldsymbol{I}_{m \times m} - \frac{\partial \boldsymbol{h}_1}{\partial \boldsymbol{x}} \boldsymbol{K}_1) \boldsymbol{x} + \boldsymbol{h}_1^{\mathrm{T}} \frac{\partial \boldsymbol{h}_1}{\partial \boldsymbol{x}_1} \boldsymbol{h}_1 - \boldsymbol{h}_1^{\mathrm{T}} \frac{\partial \boldsymbol{h}_1}{\partial \boldsymbol{x}_{2d}} \boldsymbol{K}_{v1} \left( \frac{\partial \boldsymbol{h}_1}{\partial \boldsymbol{x}_{2d}} \right)^{\mathrm{T}} \boldsymbol{h}_1$$

(32)

where $\boldsymbol{I}_{m \times m}$ is $m \times m$ dimensional identity matrix. With the Young's inequality that

$$\boldsymbol{h}_1^{\mathrm{T}} (\boldsymbol{I}_{m \times m} - \frac{\partial \boldsymbol{h}_1}{\partial \boldsymbol{x}_1} \boldsymbol{K}_1) \boldsymbol{x} \le \boldsymbol{x}^{\mathrm{T}} \boldsymbol{K}_1 \boldsymbol{x} + \frac{1}{4} \boldsymbol{h}_1^{\mathrm{T}} (\boldsymbol{I}_{m \times m} - \frac{\partial \boldsymbol{h}_1}{\partial \boldsymbol{x}_1} \boldsymbol{K}_1) \boldsymbol{K}_1^{-1} (\boldsymbol{I}_{m \times m} - \frac{\partial \boldsymbol{h}_1}{\partial \boldsymbol{x}_1} \boldsymbol{K}_1)^{\mathrm{T}} \boldsymbol{h}_1$$

$$= \boldsymbol{x}^{\mathrm{T}} \boldsymbol{K}_1 \boldsymbol{x} + \frac{1}{4} \boldsymbol{h}_1^{\mathrm{T}} \left( \boldsymbol{K}_1^{-1} - \left( \frac{\partial \boldsymbol{h}_1}{\partial \boldsymbol{x}_1} + (\frac{\partial \boldsymbol{h}_1}{\partial \boldsymbol{x}_1})^{\mathrm{T}} \right) + \frac{\partial \boldsymbol{h}_1}{\partial \boldsymbol{x}_1} \boldsymbol{K}_1 (\frac{\partial \boldsymbol{h}_1}{\partial \boldsymbol{x}_1})^{\mathrm{T}} \right) \boldsymbol{h}_1$$

(33)

$$\boldsymbol{h}_1^{\mathrm{T}} \frac{\partial \boldsymbol{h}_1}{\partial \boldsymbol{x}_1} \boldsymbol{h}_1 \le \frac{1}{2} \boldsymbol{h}_1^{\mathrm{T}} \boldsymbol{K}_1^{-1} \boldsymbol{h}_1 + \frac{1}{2} \boldsymbol{h}_1^{\mathrm{T}} \left( \frac{\partial \boldsymbol{h}_1}{\partial \boldsymbol{x}_1} \right)^{\mathrm{T}} \boldsymbol{K}_1 \frac{\partial \boldsymbol{h}_1}{\partial \boldsymbol{x}_1} \boldsymbol{h}_1$$

(34)

we may get

$$\dot{V}_{12} = \boldsymbol{h}_1^{\mathrm{T}} \left( \frac{3}{4} \boldsymbol{K}_1^{-1} - \frac{1}{4} \left( \frac{\partial \boldsymbol{h}_1}{\partial \boldsymbol{x}_1} + (\frac{\partial \boldsymbol{h}_1}{\partial \boldsymbol{x}_1})^{\mathrm{T}} \right) + \frac{3}{4} \frac{\partial \boldsymbol{h}_1}{\partial \boldsymbol{x}_1} \boldsymbol{K}_1 \left( \frac{\partial \boldsymbol{h}_1}{\partial \boldsymbol{x}_1} \right)^{\mathrm{T}} - \frac{\partial \boldsymbol{h}_1}{\partial \boldsymbol{x}_{2d}} \boldsymbol{K}_{v1} \left( \frac{\partial \boldsymbol{h}_1}{\partial \boldsymbol{x}_{2d}} \right)^{\mathrm{T}} \right) \boldsymbol{h}_1$$

(35)

with the gain condition given by Eq. (30), there is $\dot{V}_{12} < 0$.

In the first Lyapunov design of Step 2 to derive $\boldsymbol{\kappa}_2(\boldsymbol{x}_1, \boldsymbol{x}_2, \boldsymbol{x}_{2d})$, if the plant satisfies the Lipschitz condition [18]

$$\| \boldsymbol{f}_1(\boldsymbol{x}_1, \boldsymbol{x}_2) - \boldsymbol{f}_1(\boldsymbol{x}_1, \boldsymbol{x}_{2d}) \| \le L \| \boldsymbol{x}_2 - \boldsymbol{x}_{2d} \|$$

(36)

where $L$ is the Lipschitz constant, then with the condition on $\boldsymbol{K}_2$ that

$$\boldsymbol{K}_2 > \frac{L^2}{4} \left( \max \left( \mathrm{eig}(\boldsymbol{K}_1^{-1}) \right) + \max(\mathrm{eig}(\boldsymbol{M})) \right) \boldsymbol{I}_{m \times m}$$

(37)

where $\boldsymbol{M} = \left( \frac{\partial \boldsymbol{h}_1}{\partial \boldsymbol{x}_1} \right)^{\mathrm{T}} \left( (\frac{\partial \boldsymbol{h}_1}{\partial \boldsymbol{x}_{2d}})^{\mathrm{T}} \right)^{-1} \boldsymbol{K}_{v1}^{-1} \left( \frac{\partial \boldsymbol{h}_1}{\partial \boldsymbol{x}_{2d}} \right)^{-1} \frac{\partial \boldsymbol{h}_1}{\partial \boldsymbol{x}_1}$, and eig() represents the operator that solves the eigenvalue of matrix, the derived $\boldsymbol{\kappa}_2(\boldsymbol{x}_1, \boldsymbol{x}_2, \boldsymbol{x}_{2d})$ that guarantee the stability may be simplified as

$$\boldsymbol{\kappa}_2(\boldsymbol{x}_1, \boldsymbol{x}_2, \boldsymbol{x}_{2d}) = -\boldsymbol{K}_2(\boldsymbol{x}_2 - \boldsymbol{x}_{2d}) + \dot{\boldsymbol{x}}_{2d}$$

(38)

In this way Assumption 3 is also removed in deriving the controller. To verify it, construct the CLF

$$V_{21} = V_{12} + \frac{1}{2} (\boldsymbol{x}_2 - \boldsymbol{x}_{2d})^{\mathrm{T}} (\boldsymbol{x}_2 - \boldsymbol{x}_{2d})$$

(39)

Its derivative is

$$\dot{V}_{21} = \boldsymbol{x}_1^{\mathrm{T}} \dot{\boldsymbol{x}}_1 + \boldsymbol{h}_1^{\mathrm{T}} \dot{\boldsymbol{h}}_1 + (\boldsymbol{x}_2 - \boldsymbol{x}_{2d})^{\mathrm{T}} (\dot{\boldsymbol{x}}_2 - \dot{\boldsymbol{x}}_{2d})$$

$$= \boldsymbol{x}_1^{\mathrm{T}} \boldsymbol{\kappa}_1(\boldsymbol{x}_1) + \boldsymbol{h}_1^{\mathrm{T}} \boldsymbol{x}_1 + \boldsymbol{h}_1^{\mathrm{T}} \left( \frac{\partial \boldsymbol{h}_1}{\partial \boldsymbol{x}_1} \boldsymbol{f}_1(\boldsymbol{x}_1, \boldsymbol{x}_{2d}) + \frac{\partial \boldsymbol{h}_1}{\partial \boldsymbol{x}_{2d}} \dot{\boldsymbol{x}}_{2d} \right) \tag{40}$$

$$+ \left( \boldsymbol{f}_1(\boldsymbol{x}_1, \boldsymbol{x}_2) - \boldsymbol{f}_1(\boldsymbol{x}_1, \boldsymbol{x}_{2d}) \right)^{\mathrm{T}} \left( (\frac{\partial \boldsymbol{h}_1}{\partial \boldsymbol{x}_1})^{\mathrm{T}} \boldsymbol{h}_1^{\mathrm{T}} + \boldsymbol{x}_1 \right) + (\boldsymbol{x}_2 - \boldsymbol{x}_{2d})^{\mathrm{T}} (\dot{\boldsymbol{x}}_2 - \dot{\boldsymbol{x}}_{2d})$$

Presuming $\boldsymbol{\kappa}_1(\boldsymbol{x}_1) = -\boldsymbol{K}_1 \boldsymbol{x}_1$ and $\dot{\boldsymbol{x}}_{2d}$ is given by Eq. (24), then

$$\dot{V}_{21} = -\boldsymbol{x}_1^{\mathrm{T}} \boldsymbol{K}_1 \boldsymbol{x}_1 - \boldsymbol{h}_1^{\mathrm{T}} \frac{\partial \boldsymbol{h}_1}{\partial \boldsymbol{x}_{2d}} \boldsymbol{K}_{v1} \left( \frac{\partial \boldsymbol{h}_1}{\partial \boldsymbol{x}_{2d}} \right)^{\mathrm{T}} \boldsymbol{h}_1$$

$$+ \left( \boldsymbol{f}_1(\boldsymbol{x}_1, \boldsymbol{x}_2) - \boldsymbol{f}_1(\boldsymbol{x}_1, \boldsymbol{x}_{2d}) \right)^{\mathrm{T}} \left( (\frac{\partial \boldsymbol{h}_1}{\partial \boldsymbol{x}_1})^{\mathrm{T}} \boldsymbol{h}_1^{\mathrm{T}} + \boldsymbol{x}_1 \right) + (\boldsymbol{x}_2 - \boldsymbol{x}_{2d})^{\mathrm{T}} (\dot{\boldsymbol{x}}_2 - \dot{\boldsymbol{x}}_{2d}) \tag{41}$$

According to the Young's inequality, we have

$$\left( \boldsymbol{f}_1(\boldsymbol{x}_1, \boldsymbol{x}_2) - \boldsymbol{f}_1(\boldsymbol{x}_1, \boldsymbol{x}_{2d}) \right)^{\mathrm{T}} \boldsymbol{x}_1 \le \boldsymbol{x}_1^{\mathrm{T}} \boldsymbol{K}_1 \boldsymbol{x}_1 + \frac{1}{4} \left( \boldsymbol{f}_1(\boldsymbol{x}_1, \boldsymbol{x}_2) - \boldsymbol{f}_1(\boldsymbol{x}_1, \boldsymbol{x}_{2d}) \right)^{\mathrm{T}} \boldsymbol{K}_1^{-1} \left( \boldsymbol{f}_1(\boldsymbol{x}_1, \boldsymbol{x}_2) - \boldsymbol{f}_1(\boldsymbol{x}_1, \boldsymbol{x}_{2d}) \right) \tag{42}$$

$$\left( \boldsymbol{f}_1(\boldsymbol{x}_1, \boldsymbol{x}_2) - \boldsymbol{f}_1(\boldsymbol{x}_1, \boldsymbol{x}_{2d}) \right)^{\mathrm{T}} \left( \frac{\partial \boldsymbol{h}_1}{\partial \boldsymbol{x}_1} \right)^{\mathrm{T}} \boldsymbol{h}_1 \le \boldsymbol{h}_1^{\mathrm{T}} \frac{\partial \boldsymbol{h}_1}{\partial \boldsymbol{x}_{2d}} \boldsymbol{K}_{v1} \left( \frac{\partial \boldsymbol{h}_1}{\partial \boldsymbol{x}_{2d}} \right)^{\mathrm{T}} \boldsymbol{h}_1$$

$$+ \frac{1}{4} \left( \boldsymbol{f}_1(\boldsymbol{x}_1, \boldsymbol{x}_2) - \boldsymbol{f}_1(\boldsymbol{x}_1, \boldsymbol{x}_{2d}) \right)^{\mathrm{T}} \boldsymbol{M} \left( \boldsymbol{f}_1(\boldsymbol{x}_1, \boldsymbol{x}_2) - \boldsymbol{f}_1(\boldsymbol{x}_1, \boldsymbol{x}_{2d}) \right) \tag{43}$$

Further with the Lipschitz condition (36) and the gain condition (37), it will be found that $\dot{V}_{21} < 0$. In addition, we may also use the first-order approximation of $\boldsymbol{f}_1(\boldsymbol{x}_1, \boldsymbol{x}_2) - \boldsymbol{f}_1(\boldsymbol{x}_1, \boldsymbol{x}_{2d})$ in Eq. (41) to derive

$$\boldsymbol{\kappa}_2(\boldsymbol{x}_1, \boldsymbol{x}_2, \boldsymbol{x}_{2d}) = -\boldsymbol{K}_2(\boldsymbol{x}_2 - \boldsymbol{x}_{2d}) - \left( \frac{\partial \boldsymbol{h}_1}{\partial \boldsymbol{x}_{2d}} \right)^{\mathrm{T}} \left( (\frac{\partial \boldsymbol{h}_1}{\partial \boldsymbol{x}_1})^{\mathrm{T}} \boldsymbol{h}_1 + \boldsymbol{x}_1 \right) + \dot{\boldsymbol{x}}_{2d} \tag{44}$$

which is more complex but may require a smaller $\boldsymbol{K}_2$.

In general, the larger the control gains are, the simpler the form of the controller might be. With these simplifications during the design, the form of final control law may be simplified.

### 4.2 System containing strict-feedback dynamics

We have considered the problem where analytic expressions of all the (virtual) controls are not available, yet the design method proposed is also applicable to include the situation where some explicit expression of the (virtual) control may be obtained. Variation of such case may be abundant, but the key is to employ the dynamic backstepping whenever it is necessary. For example, consider plant of the form

$$\dot{\boldsymbol{x}}_1 = \boldsymbol{f}_1(\boldsymbol{x}_1, \boldsymbol{x}_2) \tag{45}$$

$$\dot{\boldsymbol{x}}_2 = \boldsymbol{f}_2(\boldsymbol{x}_1, \boldsymbol{x}_2) + \boldsymbol{g}_2(\boldsymbol{x}_1, \boldsymbol{x}_2) \boldsymbol{u} \tag{46}$$

This plant contains a strict-feedback dynamics where the actual control $\boldsymbol{u}$ may be directly derived. In the design procedure, Step 1 is same as that in Sec. 3.3, and it contains two Lyapunov designs. In Step 2, the CLF is same as Eq. (25) and through the algebraic control equation

$$\boldsymbol{f}_2(\boldsymbol{x}_1, \boldsymbol{x}_2) + \boldsymbol{g}_2(\boldsymbol{x}_1, \boldsymbol{x}_2)\boldsymbol{u} = \boldsymbol{\kappa}_2(\boldsymbol{x}_1, \boldsymbol{x}_2, \boldsymbol{x}_{2d}) \tag{47}$$

we may directly get the control law

$$\boldsymbol{u} = \boldsymbol{g}_2^{-1}\left(-\boldsymbol{f}_2(\boldsymbol{x}_1, \boldsymbol{x}_2) + \boldsymbol{\kappa}_2(\boldsymbol{x}_1, \boldsymbol{x}_2, \boldsymbol{x}_{2d})\right) \tag{48}$$

where $\boldsymbol{\kappa}_2(\boldsymbol{x}_1, \boldsymbol{x}_2, \boldsymbol{x}_{2d})$ is same to that in Eq (14). In this step, only one Lypunov design is required. The proof for the closed-loop stability is similar to that in Sec. 3.2, by nullifying the term regarding the implicit algebraic control equation arising from the second step.

## 4.3 One treatable singularity case

Before giving Theorem 1, we presented Assumption 2 as a precondition. For one special case where Assumption 2 fails, the method is also applicable. Consider a scalar illustrative example

$$\dot{x}_1 = x_1(x_1 + u + u^3) \tag{49}$$

where $x_1$ is the state and $u$ is the control. The algebraic control equation may be in the following form as

$$h = x_1(x_1 + u + u^3) + K_1 x_1 \tag{50}$$

where $K_1$ is a scalar positive control gain. Apparently, Assumption 2 is not satisfied at the origin and there are infinite solutions of $u$ when $x_1 = 0$. However, for such case, a continuous mapping $u^* = C(x_1)$ determined by Eq. (50) maybe exist. Actually this problem is solvable, by transforming the former algebraic control equation to

$$\tilde{h} = \frac{1}{x_1}h = (x_1 + u + u^3) + K_1 \tag{51}$$

then a definite control under the dynamic backstepping frame may be determined.

Now we consider a general case of the $i$-th level dynamics of the cascade pure-feedback system

$$\dot{x}_i = \boldsymbol{f}_i(\boldsymbol{x}_1, ..., \boldsymbol{x}_i, \boldsymbol{x}_{i+1}) \tag{52}$$

For the right-hand function, if the virtual control $\boldsymbol{x}_{i+1}$ may take arbitrary value that the following equation holds

$$\boldsymbol{f}_i(\boldsymbol{0}, ..., \boldsymbol{0}, \boldsymbol{x}_{i+1}) = \boldsymbol{0} \tag{53}$$

then to avoid singularity, the algebraic control equation may be formulate as

$$\tilde{\boldsymbol{h}}_i = \boldsymbol{R}(\boldsymbol{x}_1, ..., \boldsymbol{x}_i)\boldsymbol{h}_i = \boldsymbol{0} \tag{54}$$

which satisfies that $\dfrac{\partial \tilde{\boldsymbol{h}}_i}{\partial \boldsymbol{x}_{(i+1)d}} \neq \boldsymbol{0}$ at the controlled domain $\mathbb{D}$ and where $\boldsymbol{h}_i = \boldsymbol{f}_i(\boldsymbol{x}_1,...,\boldsymbol{x}_i,\boldsymbol{x}_{(i+1)d}) - \boldsymbol{\kappa}_i(\boldsymbol{x}_1,...,\boldsymbol{x}_i,\boldsymbol{x}_{2d},...,\boldsymbol{x}_{id})$ is the original implicit equation, and $\boldsymbol{R}(\boldsymbol{x}_1,...,\boldsymbol{x}_i)$ is a matrix-valued function. Note that for the virtual control $\boldsymbol{x}_{(i+1)d}$, it should be guaranteed that the mapping $\boldsymbol{x}_{(i+1)d} = \boldsymbol{C}_i(\boldsymbol{x}_1,...,\boldsymbol{x}_i,\boldsymbol{x}_{2d},...,\boldsymbol{x}_{id})$ implicitly determined by Eq. (54) is continuous and satisfies $\boldsymbol{C}_i(\boldsymbol{0},...,\boldsymbol{0}) = \boldsymbol{0}$ through setting $\boldsymbol{\kappa}_i(\boldsymbol{x}_1,...,\boldsymbol{x}_i,\boldsymbol{x}_{2d},...,\boldsymbol{x}_{id})$. The modification will be employed in the Lyapunov design to deduce the control law as

$$V_{i2} = V_{i1} + \frac{1}{2}\tilde{\boldsymbol{h}}_i^{\mathrm{T}}\tilde{\boldsymbol{h}}_i \tag{55}$$

and

$$V_{(i+1)1} = V_{i2} + \frac{1}{2}\left(\tilde{\boldsymbol{f}}_i(\boldsymbol{x}_1,...,\boldsymbol{x}_i,\boldsymbol{x}_{i+1}) - \tilde{\boldsymbol{f}}_i(\boldsymbol{x}_1,...,\boldsymbol{x}_i,\boldsymbol{x}_{(i+1)d})\right)^{\mathrm{T}}\left(\tilde{\boldsymbol{f}}_i(\boldsymbol{x}_1,...,\boldsymbol{x}_i,\boldsymbol{x}_{i+1}) - \tilde{\boldsymbol{f}}_i(\boldsymbol{x}_1,...,\boldsymbol{x}_i,\boldsymbol{x}_{(i+1)d})\right) \tag{56}$$

where the modified term $\tilde{\boldsymbol{f}}_i(\boldsymbol{x}_1,...,\boldsymbol{x}_i,\boldsymbol{x}_{i+1}) = \boldsymbol{R}(\boldsymbol{x}_1,...,\boldsymbol{x}_i)\boldsymbol{f}_i(\boldsymbol{x}_1,...,\boldsymbol{x}_i,\boldsymbol{x}_{i+1})$ aims to guarantee the validity of Assumption 3, i.e., $\tilde{\boldsymbol{f}}_i(\boldsymbol{x}_1,...,\boldsymbol{x}_i,\boldsymbol{a}) \neq \tilde{\boldsymbol{f}}_i(\boldsymbol{x}_1,...,\boldsymbol{x}_i,\boldsymbol{b})$ when $\boldsymbol{a} \neq \boldsymbol{b}$.

### 4.4 Tracking problem

In the preceding we focused on the stabilization problem, and the proposed approach is also applicable to the tracking problem. For the pure-feedback cascade plant given by Eqs. (8) and (9), presume that the reference signal is $\boldsymbol{r}$ and the output function is

$$\boldsymbol{y} = \boldsymbol{x}_1 \tag{57}$$

with extra consideration about the effect of reference signal, the design procedure that achieves asymptotically tracking is similar to the stability controller design. The CLF constructed during the design procedure is

$$V_{11} = \frac{1}{2}(\boldsymbol{x}_1 - \boldsymbol{r})^{\mathrm{T}}(\boldsymbol{x}_1 - \boldsymbol{r}) \tag{58}$$

and the resulting (virtual) control, i.e., $\boldsymbol{x}_{2d}$ and $\boldsymbol{u}$, are similar to the stabilization controller but includes the dynamics of $\boldsymbol{r}$. Similarly, $\lim_{t \to \infty}(\boldsymbol{y} - \boldsymbol{r}) = 0$ are guaranteed through the Lyapunov principle.

## 5. Illustrative examples

### 5.1 Example 1: Stabilization of pure-feedback system

A nonlinear system in the pure-feedback form from Refs. [14] and [15] is considered

$$\dot{x}_1 = x_1 + x_2 + \frac{x_2^3}{5}$$

$$\dot{x}_2 = x_1 x_2 + u + \frac{u^3}{7}$$

(59)

where $x_1$, $x_2$ are the states and $u$ is the control. The control objective is to stabilize the states to the origin. According to the dynamic backstepping controller design procedure in Sec. 3.3, the implicit nonlinear algebraic control equations are

$$h_1 = x_1 + x_{2d} + \frac{x_{2d}^3}{5} + K_1 x_1$$

(60)

$$h_2 = x_1 x_2 + u + \frac{u^3}{7} + K_2 \left(1 + 0.6 x_2^2\right) \left(\left(x_2 + 0.2 x_2^3\right) - \left(x_{2d} + 0.2 x_{2d}^3\right)\right)$$

$$+ \left(\frac{1}{1 + 0.6 x_2^2}\right) \left(\left(2 - K_1 - K_1^2\right) x_1 + 2(1 + K_1) h_1 + K_{v1} \left(1 + 0.6 x_{2d}^2\right)^2 h_1\right)$$

(61)

The control gains (in scalar form) $K_1$, $K_2$, $K_{v1}$ and $K_{v2}$ are all set to be 1. The initial states of the system are arbitrarily set to be $\begin{bmatrix} x_1 \\ x_2 \end{bmatrix}\bigg|_{t=0} = \begin{bmatrix} 0.5 \\ 0 \end{bmatrix}$, and the initial conditions for the augmented dynamics are $x_{2d}|_{t=0} = 0$ and $u|_{t=0} = 0$. The simulation results about the states are plotted in Fig. 1. It is shown that the stats are stabilized to the origin as expected. The virtual control $x_{2d}$ and control $u$, which are augmented states in the controller, are plotted in Fig. 2. The profiles about the implicit algebraic control equations, i.e., $h_1$ and $h_2$, are presented in Fig. 3, showing that they approaches zero rapidly.

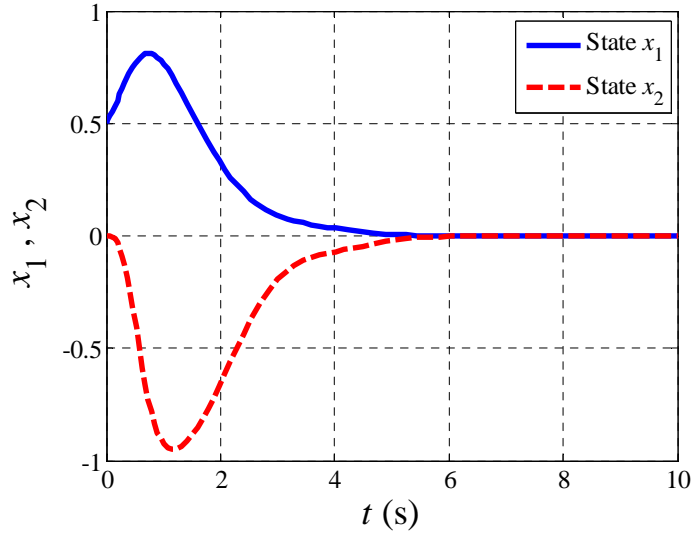

Fig. 1 The states profiles in Example 1

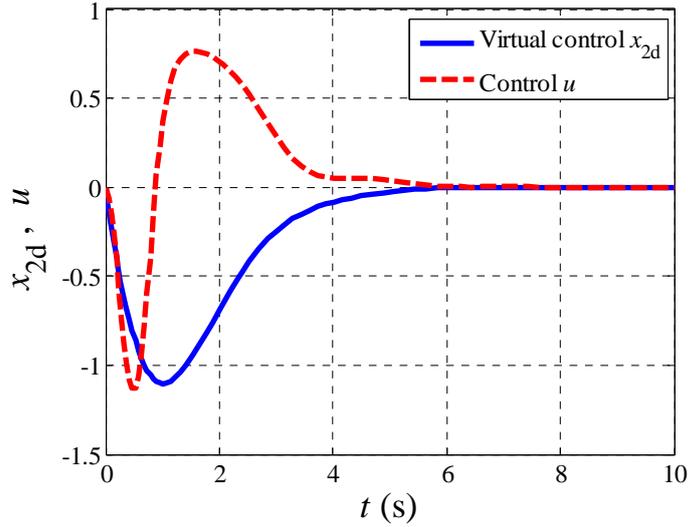

Fig. 2 The virtual control and control profiles in Example 1

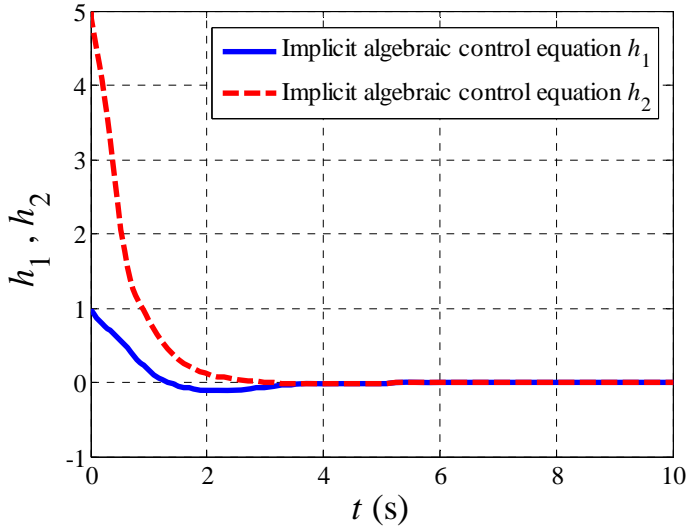

Fig .3 The implicit algebraic control equation profiles in Example 1

## 5.2 Example 2: Stabilization of system containing strict-feedback dynamics

A simplified Jet engine model from Ref. [1], which includes the strict-feedback dynamics, are considered. The dynamic equations are

$$\dot{R} = -\sigma R^2 - \sigma R(2\phi + \phi^2)$$
$$\dot{\phi} = -\frac{3}{2}\phi^2 - \frac{1}{2}\phi^3 - 3R\phi - 3R - \psi \tag{62}$$
$$\dot{\psi} = -u$$

where $R$, $\phi$, $\psi$ are the states and $u$ is the control. For this plant, the dynamics regarding $R$ is in the pure-feedback form and the dynamics regarding $\phi$ and $\psi$ are in the strict-feedback form. Thus within the proposed controller design scheme, it requires three steps and the first step includes two Lyapunov designs.

In Ref. [1], the controller is artfully designed by using the input-to-state stability of dynamics regarding $R$. Here it is addressed under a unified frame. However, this work does not aim to show the superiority in performance of the proposed method. It is just used to indicate its capacity to deal with a model including the pure-feedback dynamics.

Especially in designing the controller, $\kappa_1(R)$ is set to be $\kappa_1(R) = -K_1 R^3$ for the requirement that the augmented state $\phi_d$ continuously approaches zero when $R$ approaches zero. Also, the treatable singularity is avoided by employing a modified implicit algebraic control equation as

$$\tilde{h}_1 = \frac{1}{R} h_1 = -\sigma R - \sigma(2\phi + \phi^2) + K_1 R^2 \qquad (63)$$

The scalar control gains $K_1$, $K_2$, $K_3$ and $K_{v1}$ are all set to be 1. The initial states of the system are arbitrarily set to be $\begin{bmatrix} R \\ \phi \\ \psi \end{bmatrix}\Big|_{t=0} = \begin{bmatrix} 2 \\ 5 \\ -5 \end{bmatrix}$, and the initial condition for the augmented dynamics is $\phi_d|_{t=0} = 0$.

Figure 4 gives the profiles of the states, and all of them are stabilized to the origin. The profiles regarding virtual control $\phi_d$ (a augmented state) and $\psi_d$ (function of $R$, $\phi$, and $\phi_d$), and control $u$ (function of $R$, $\phi$, $\psi$, and $\phi_d$) are presented in Fig. 5. At the initial time, the values of $\psi_d$ and $u$ are large while $\phi_d$ increases fast from zero, the initial condition prescribed. For the implicit nonlinear equation $\tilde{h}_1$, its profile is given in Fig. 6 and it approaches zero as expected.

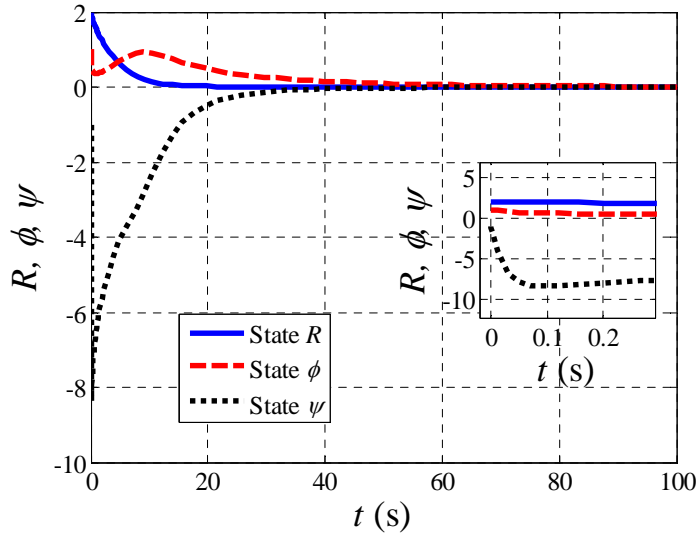

Fig. 4 The states profiles in Example 2

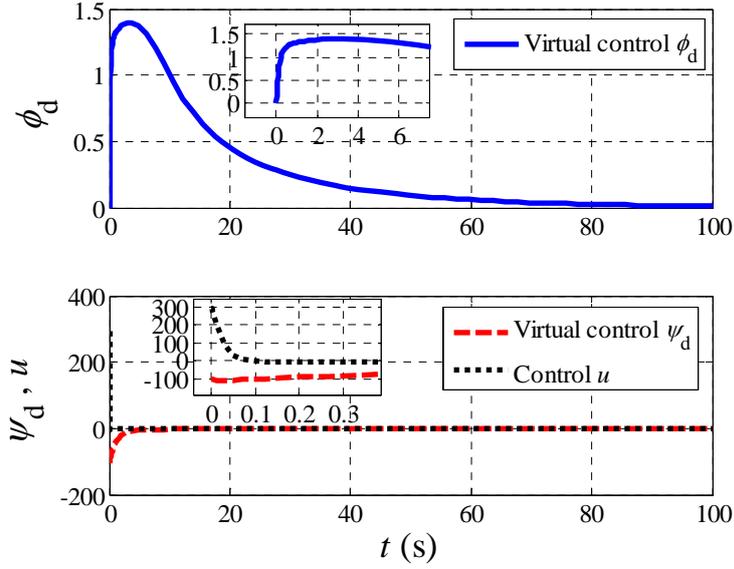

Fig. 5 The virtual control and control profiles in Example 2

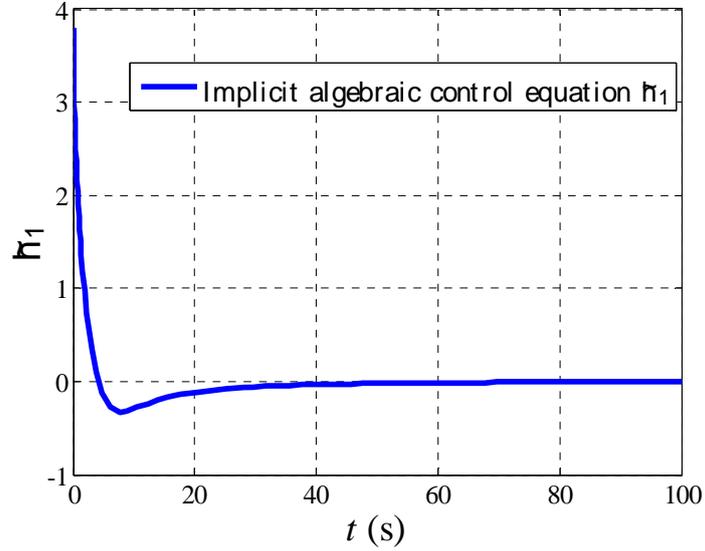

Fig .6 The implicit algebraic control equation profile in Example 2

## 5.3 Example 3: Signal tracking

Again consider the pure-feedback model given in Example 1. The tracking problem is defined in Refs. [14] and [15], with the famous van der Pol oscillator taken as the reference model.

$$\frac{\mathrm{d}}{\mathrm{d}t}\begin{bmatrix} r \\ \dot{r} \end{bmatrix} = \begin{bmatrix} \dot{r} \\ -r + 0.2(1 - r^2)\dot{r} \end{bmatrix} \tag{64}$$

This model yields a limit cycle trajectory for any initial states except $\begin{bmatrix} r \\ \dot{r} \end{bmatrix} = \begin{bmatrix} 0 \\ 0 \end{bmatrix}$. The output of the controlled system is $x_1$, and the control objective is to make $x_1$ follow the reference signal $r$. The dynamic backstepping controller is designed similarly as Example 1, along with extra consideration on the reference signal. Now the implicit nonlinear algebraic control equations are

$$h_1 = x_1 + x_{2d} + \frac{x_{2d}^3}{5} + K_1(x_1 - r) - \dot{r} \tag{65}$$

$$h_2 = x_1 x_2 + u + \frac{u^3}{7} + K_2\left(1 + 0.6 x_2^{\,2}\right)\left(\left(x_2 + 0.2 x_2^{\,3}\right) - \left(x_{2d} + 0.2 x_{2d}^{\,3}\right)\right)$$
$$+ \left(\frac{1}{1 + 0.6 x_2^{\,2}}\right)\left(\left(2 - K_1 - K_1^{\,2}\right)x_1 + 2(1 + K_1)h_1 + K_{v1}\left(1 + 0.6 x_{2d}^{\,2}\right)^2 h_1 - 2r - K_1\dot{r} - \ddot{r}\right) \tag{66}$$

The control gains $K_1$, $K_2$, $K_{v1}$ and $K_{v2}$ are all again set to be 1. The initial conditions for the reference model and the controlled system are $\begin{bmatrix} r \\ \dot{r} \end{bmatrix}\bigg|_{t=0} = \begin{bmatrix} 0.5 \\ 0 \end{bmatrix}$ and $\begin{bmatrix} x_1 \\ x_2 \end{bmatrix}\bigg|_{t=0} = \begin{bmatrix} 0.5 \\ 0 \end{bmatrix}$, respectively. The initial conditions for the augmented dynamics are also $x_{2d}\big|_{t=0} = 0$ and $u\big|_{t=0} = 0$. Figure 7 plots the output trajectory in the tracking and the tracking error, showing that the nearly exact tracking performance is achieved. The periodical virtual control $x_{2d}$ and control $u$ are presented in Fig. 8. Figure 9 gives the profiles of $h_1$ and $h_2$. They also approaches zero quickly.

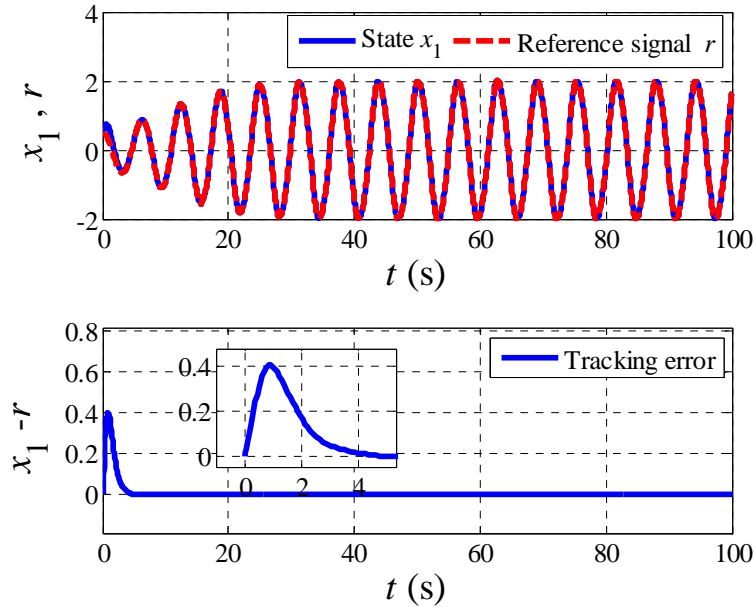

Fig. 7 The tracking results to the reference signal in Example 3

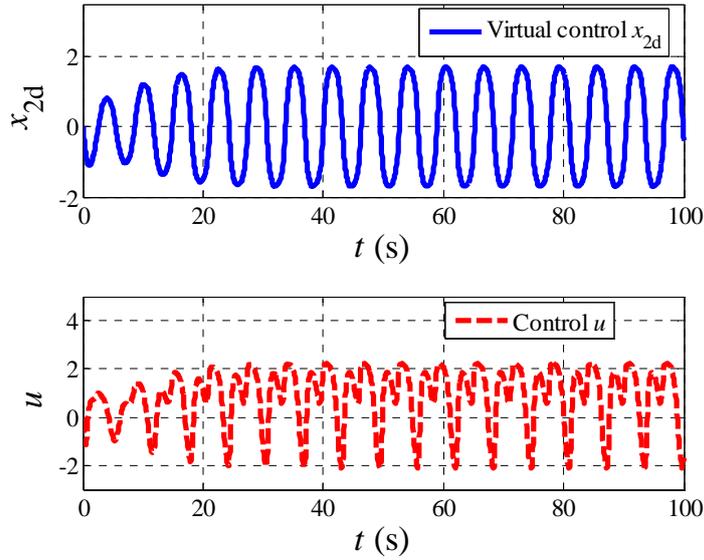

Fig. 8 The virtual control and control profiles in Example 3

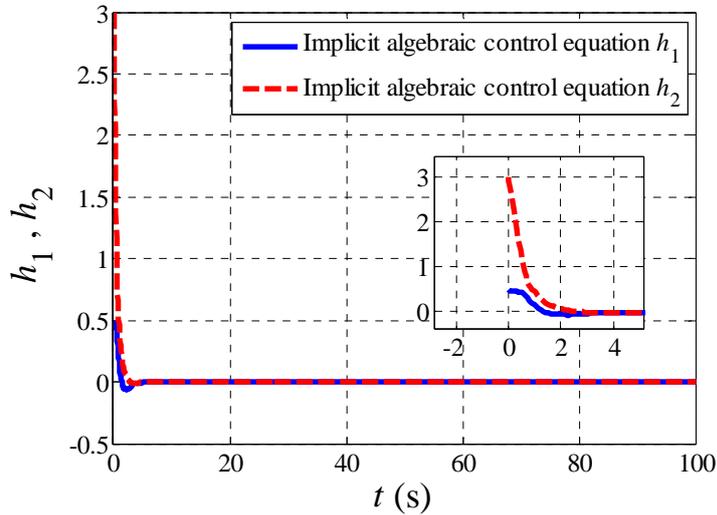

Fig .9 The implicit algebraic control equation profiles in Example 3

## 6. Conclusion

A general backstepping controller design frame is developed for the pure-feedback system. The idea is to introduce new dynamics to describe the (virtual) control and it solves the implicit nonlinear algebraic equation in an asymptotically way, from a control-based view. Situations where controller may be simplified are discussed, which will alleviate the problem of "explosion of terms". This paper provides the solution for the general pure-feedback system controller design problem with exact model, and it may be extended to address the problems with uncertainties, as the adaptive backstepping method studied on the strict-feedback systems. Moreover, consideration on the control saturation and time delay may also be investigated. These problems will be studied in the future to further complete this method.